
\input harvmac

\def\pf{{\rm Pf ~}}
\font\zfont = cmss10 
\font\litfont = cmr6

\def\bigone{\hbox{1\kern -.23em {\rm l}}}
\def\ZZ{\hbox{\zfont Z\kern-.4emZ}}
\def\half{{\litfont {1 \over 2}}}

\def\gG{{\cal G}}

\def\CM{{\cal M}}

\def\bZ{{\bf Z}}
\def\Re{{\rm Re ~}}
\def\Im{{\rm Im ~}}

\def\lfm#1{\medskip\noindent\item{#1}}
\noblackbox
\def\cl{{\cal L}}

\def\pf{{\rm Pf ~}}
\def\hH{{\cal H}}
\def\np#1#2#3{Nucl. Phys. B{#1} (#2) #3}
\def\pl#1#2#3{Phys. Lett. {#1}B (#2) #3}

\def\physrev#1#2#3{Phys. Rev. {D#1} (#2) #3}

\def\prep#1#2#3{Phys. Rep. {#1} (#2) #3}

\def\ev#1{\langle#1\rangle}
\def\tilde{\widetilde}
\font\litfont = cmr6
\def\half{{\litfont {1 \over 2}}}
\def\mmp{\tilde E^{(i)}E^{(i)}}

\Title{hep-th/9408155, RU-94-68, IASSNS-HEP-94/67}
{\vbox{\centerline{Phases of $N=1$ supersymmetric gauge theories }
\smallskip
\centerline{in four dimensions}}}
\bigskip
\centerline{K. Intriligator$^1$ and N. Seiberg$^{1,2}$}
\vglue .5cm
\centerline{$^1$Department of Physics and Astronomy}
\centerline{Rutgers University}
\centerline{Piscataway, NJ 08855-0849, USA}
\vglue .3cm
\centerline{$^2$ School of Natural Sciences}
\centerline{Institute for Advanced Study}
\centerline{Princeton, NJ 08540, USA}

\medskip

\noindent
We exhibit $N=1$ supersymmetric field theories in confining, Coulomb
and Higgs phases.  The superpotential and the gauge kinetic terms are
holomorphic and can be determined exactly in the various phases.  The
Coulomb phase generically has points with massless monopoles.  When
they condense, the theory undergoes a phase transition to a confining
phase.  When there are points in the Coulomb phase with massless
electric charges, their condensation leads to a transition to a Higgs
phase.  When the Higgs and confinement phases are distinct, we expect
to find massless interacting gluons at the transition point between
them.

\Date{8/94}

\newsec{Introduction}

Gauge theories in four dimensions can manifest themselves in three known
phases: Coulomb, Higgs, and confinement.  As the parameters $g_I$ of the
theory are varied, a phase transition between them can take
place.  In supersymmetric theories there is a new phenomenon.  When
supersymmetry is unbroken the theory can have several inequivalent
ground states for fixed values of the parameters.  These different
ground states can be in different phases.  In many cases they form a
continuous manifold -- a moduli space of vacua
\ref\nati{N. Seiberg, hep-th/9402044, \physrev{49}{1994}{6857}.}
-- parametrized by massless fields (moduli), $\Phi^r$.  There can then
also be phase transitions in the space of vacua.  Therefore, we will be
interested in the total space whose base is the parameter space and the
moduli space is fibered over it. (Note that for different values of the
parameters the fiber can have different topology and even different
dimension.)

A crucial element in our analysis is holomorphy.  The superpotential
$W$ and the coefficient $\tau$ of the gauge kinetic terms are
holomorphic in the moduli fields $\Phi^r$.  They must also be
holomorphic in all the coupling constants $g_I$
\ref\nonren{N. Seiberg, hep-ph/9309335, \pl{318}{1993}{469}.}.
This holomorphy constrains the kind of transitions which are possible.
First, there cannot be any first order transitions between different
phases.  This is because the energy of every ground state is zero when
supersymmetry is unbroken and, hence, there are no level crossings of
vacua.  Furthermore, because of holomorphy, phase boundaries are always
of real codimension two or larger.  The known examples are of two
classes:
\lfm{1.} Most of the moduli space is in one phase and it has a small
subspace of complex codimension larger or equal to one of another phase
\nati.
\lfm{2.} The moduli space has separate branches in different phases
which touch each other at transition points
\nref\ntwoi{N. Seiberg and E. Witten, RU-94-52, IASSNS-HEP-94/43,
hep-th/9407087.}%
\nref\ntwoii{N. Seiberg and E. Witten, RU-94-60, IASSNS-HEP-94/55,
hep-th/9408099.}%
\def\ntwo{\refs{\ntwoi,\ntwoii}}%
\ntwo.

\noindent
Below we will see more examples in both classes.

\nref\cerne{G. Veneziano and S. Yankielowicz, \pl{113}{1982}{321};
T.R. Taylor, G. Veneziano, and S. Yankielowicz,
\np{218}{1983}{493}.}%
\nref\ads{I. Affleck, M. Dine, and N. Seiberg, \np{241}{1984}{493};
\np{256}{1985}{557}.}%
\nref\nsvz{V.A. Novikov, M.A. Shifman, A. I.  Vainstain and V. I.
Zakharov, \np{223}{1983}{445}; \np{260}{1985}{157}}%
\nref\cernrev{D. Amati, K. Konishi, Y. Meurice, G. C. Rossi, and G.
Veneziano, \prep{162}{1988}{169} and references therein.}%
\nref\kl{V. Kaplunovsky and J. Louis, hep-th/9402005, UTTG-94-1,
LMU-TPW-94-1.}%
\nref\iin{K. Intriligator, R.G. Leigh and N. Seiberg, hep-th/9403198,
\physrev{50}{1994}{1092}; K. Intriligator, RU-94-57, hep-th/9407106,
Phys. Lett. B in press}%

Most of the analysis of supersymmetric gauge theories has been devoted
to theories with matter fields in the fundamental representation
\refs{\cerne-\iin,\nati}, where there is no distinction between the
confining and the Higgs phase
\ref\fund{T. Banks and E. Rabinovici, \np{160}{1979}{349}; E. Fradkin
and S. Shenker, \physrev{19}{1979}{3682}.}.
These theories did not have a manifold of ground states in the Coulomb
phase.  Other theories (like those of \ntwo\ and the ones below) do have
a moduli space of vacua in the Coulomb phase.  Trying to apply the
techniques of \iin\ to these theories, one finds superpotentials which
seem to miss the Coulomb phase.  Below we will interpret these
superpotentials as giving accurate descriptions of the confining phases
of these theories.

One of the main points of this paper is an extension of some of the work
of \ntwo\ from $N=2$ to $N=1$ theories.  In the second section we
discuss the Coulomb phase of $N=1$ theories.  We show that many
of the phenomena in \ntwo\ are also present for generic $N=1$ supersymmetric
theories which have a Coulomb phase.  In particular, we will argue that there
are points on the moduli space with massless magnetic monopoles.  With a
suitable perturbation these monopoles condense and the theory passes to
a confining phase.

In the third section we combine the techniques of \iin\ and those
developed in \ntwo\ and in section 2 to analyze illustrative examples. The
confining phase is well described by the superpotentials of \iin.
These are not valid in the Coulomb phase or in a Higgs phase when it
is distinct from the confining phase.  The reason is that new degrees
of freedom should be included for a proper description of the
transition point.  It has already been observed in \ntwo, and we will
show more generally in
$N=1$ examples below, that sometimes there is no Lagrangian
which describes the low energy physics everywhere on the moduli space.
We must be content with effective Lagrangians which describe only
patches of the moduli space.  In the overlap regions between different
patches the different Lagrangians describe the same massless modes but
include different massive modes.

\newsec{The Coulomb phase}

A gauge theory in the Coulomb phase has a massless photon and therefore it
is subject to standard electric-magnetic duality.  The gauge kinetic
term in  the low energy effective Lagrangian is
\eqn\nzerolag{{1\over 64\pi} \Im [\tau(g_I, \Phi^r) (F+i{}^*F)^2]={1\over
32\pi} (\Im \tau F^2 +
\Re \tau F{}^*F)}
where $\tau$ gives the effective coupling $\tau = {\theta_{eff} \over
\pi}+i{8\pi \over g_{eff}^2}$, $g_I$ are the coupling constants of the
underlying microscopic theory, and $\Phi^r$ are some light fields.
(We are here normalizing $\tau$ as in \ntwoii\ because we will
consider examples with matter fields in the fundamental of $SU(2)$.)
Under the electric-magnetic duality transformation $S$, the term
\nzerolag\ is mapped into
\eqn\nzerolagd{{1\over 64\pi} \Im [\tau_d (F_d+i{}^*F_d)^2]}
where $F_d$ is the dual of $F$ and $\tau_d=-1/\tau$.  The effect on the
spectrum is to take states with magnetic and electric charges
$(n_m,n_e)$ to $(n_e, -n_m)$.  There is another duality transformation,
denoted by $T$, which maps $\tau \rightarrow \tau+1$ and has the effect
of shifting $\theta$.  As in
\ref\wittene{E. Witten, \pl{86}{1979}{283}.},
this changes the electric charges of states with magnetic charge as
$T:(n_m, n_e)\rightarrow (n_m, n_e+n_m)$.  Together, these two
transformations generate the infinite duality group $SL(2,\bZ)$ (only
$PSL(2,\bZ)$ acts on $\tau$).

The function $\tau(g_I, \Phi^r)$ in \nzerolag\ is not necessarily
single valued.  For example, as a microscopic $\theta$ parameter is
shifted by $2\pi$, $\Re\tau$ is shifted by an integer.  More
generally, changing $g_I$ and $\Phi^r$ along a closed path can
transform $\tau$ and the spectrum by an $SL(2,\bZ)$ transformation.
Therefore, $\tau$ is not a function but a section of an $SL(2,\bZ)$
bundle over the space of $g_I$ and $\Phi^r$.  This was observed in
\ntwo\ in an $N=2$ supersymmetric context but is clearly more general.
In non-supersymmetric theories it is not easy to determine
$\tau(g_I,\Phi^r)$.  However, as we will see below, $\tau(g_I,\Phi^r)$
can often be found exactly in $N=1$ supersymmetric theories.

We consider $N=1$ supersymmetric gauge theories based on gauge group
$\gG$ (taken to be semi-simple) with matter fields $\phi_i$ in
representations $R_i$ of $\gG$.  Typically the potential has flat
directions with nonzero $\phi _i$ expectation values.  These break
$\gG$ to a subgroup $\hH$.  This moduli space of classical ground states
is labeled by the expectation values of various gauge invariant
polynomials $\Phi^r$ of the $\phi_i$.  We will be interested in an
effective low energy theory with the massive $\gG /\hH$ vector bosons
integrated out.  If the remaining $\hH$ super Yang-Mills theory is
non-abelian, it will dynamically generate a mass gap, confine, and the
light fields in the low energy effective theory are simply the gauge
singlets $\Phi^r$.  However, when $\hH$ contains an abelian factor, say
a single $U(1)$, the spectrum includes a massless photon supermultiplet
and the theory is in the Coulomb phase.  The effective Lagrangian in
this phase has a term which is the $N=1$ supersymmetric version of
\nzerolag\
\eqn\photonl{\cl=\dots +{1\over 16\pi}Im\int d^2 \theta\tau (g_I,\Phi^r)
W_{\alpha}W^{\alpha},}
where $W_{\alpha}$ is the photon field strength superfield.

Supersymmetry requires $\tau$ in \photonl\ to be holomorphic\foot{It
is important here that we are discussing a Wilsonian effective action
\ref\russ{M.A. Shifman and A.I Vainshtein, \np{277}{1986}{456};
\np{359}{1991}{571}.}.}
in the chiral superfields $\Phi^r$ and also in all the coupling
constants $g_I$ \nonren.  This holomorphy often enables us to determine
$\tau$ exactly.  This was done in \ntwo\ in some $N=2$ theories and will
be done here for some $N=1$ theories.  Although $\tau$ is a holomorphic
function of its arguments, it is not single valued.  Exactly as with
\nzerolag , there is an electric-magnetic duality transformation which
maps $W\rightarrow W_D$ and $\tau\rightarrow \tau _D=-1/\tau$; in
addition there is the operation $T$ of shifting the $\theta$ angle by
$2\pi$ (or $\pi$).  Once again, $\tau$ is a section of an $SL(2,\bZ)$
bundle over the space of $\Phi^r$ and $g_I$.

For simplicity, we will consider the case where there is a single $\tau$
which depends only on a single light field $U$.  For large $U$ the
underlying $\gG $ gauge theory is weakly coupled and the one loop beta
function in the microscopic theory leads to
\eqn\asymptau{\tau \approx {i M \over 2\pi} \log {U\over \Lambda^b}}
for some integers $M$ and $b$.  $\Lambda$ is the scale of $\gG$, which
we assume here is simple (below we will also consider a semi-simple
$\gG$ example).  As we circle around infinity, $U \rightarrow e^{2\pi
i} U$, $\tau \rightarrow \tau - M$; i.e.\ $\tau$ is transformed by
$\CM_\infty=T^{-M}$, which leaves the low energy effective gauge
coupling ${1 \over g_{eff}^2} \sim \Im \tau$ unchanged.  The following
argument shows that $\Im \tau$ cannot be single valued in the interior
of the moduli space \ntwo.  If $\Im \tau$ is single valued, it is a
harmonic function, which cannot be positive definite.  There would
then be regions in the moduli space where $g_{eff}$ is imaginary.
This unphysical conclusion can be avoided if the topology of the
moduli space is complicated in the interior or, as found in \ntwo,
there are several (at least two) singular values $U_i$ of $U$ with
monodromies $\CM_i$ around them which do not commute with $\CM_\infty=
T^{-M}$.

The monodromies $\CM_i$ around the $U_i$ must have a physical
interpretation.  The simplest one is that they are associated with $k_i$
massless particles at the singularity.  The low energy superpotential
near $U_i$ then has the form
\eqn\lowengw{W_L{}^{(i)} =  (U-U_i) \sum_{l=1}^{k_i} c_l^{(i)} \tilde
E_l^{(i)} E_l^{(i)} + \CO ((U-U_i)^2)}
where $\tilde E_l^{(i)}$ and $E_l^{(i)}$ are the new massless states.
If the constants $c_l^{(i)}$ are nonzero, these states acquire a mass
of order $ \CO ((U-U_i)) $ away from the singularity.  Therefore, the
one loop beta function in the low energy theory leads to
\eqn\lowbet{\tau_i \approx -{ i k_i \over 2 \pi} \log (U-U_i) }
(we assume for simplicity that, as in \ntwo, all the $E_l^{(i)}$ have
charge one; the generalization to other cases is straightforward)
where $\tau_i$ is the coupling to the low energy photon.  $\tau_i$ is
related to $\tau$ in the asymptotic region by a duality transformation
$N_i$.  It is clear from \lowbet\ that the monodromy in $\tau_i$ is
$T^{k_i}$.  Therefore, the monodromy in $\tau$ is
\eqn\monti{\CM_i = N_i^{-1} T^{k_i} N_i .}
For $\CM_i$ to not commute with $\CM_\infty= T^{-M}$, the transformation
$N_i$ must be non-trivial.  This means that the massless particles
$E_l^{(i)}$ at $U_i$ are magnetically charged.

As discussed in \ntwo , because $\tau$ is a section of an
$SL(2,\bZ)$ bundle it is naturally interpreted as the modular parameter
of a torus.  A torus is conveniently described by the one complex
dimensional curve in $C^2$:
\eqn\torusc{y^2=x^3+ax^2+bx+c,}
where $(x,y)\in C^2$ and $a$, $b$ and $c$ are parameters to be related
to $U$ and the various coupling constants and scales.  The function
$\tau$ is singular when the torus is singular, which is when
\eqn\toruss{x^3+ax^2+bx+c=0\qquad\hbox{and}\qquad 3x^2+2ax+b=0.}
Eliminating $x$, this is when the discriminant of the cubic equation in
\toruss\ vanishes: $\Delta (a,b,c)=0$ where
\eqn\singabc{\Delta=4a^3c-b^2a^2-18abc+4b^3+27 c^2.}
As discussed in \ntwo, the order of the zero can be used to determine
the monodromy around the singularity.

Important constraints \ntwoii\ on the dependence of the coefficients in
\torusc\ on $U$ and the coupling constants are the following:
\lfm{1.} In the weak coupling limit $\Lambda =0$ the curve should be
singular for every $U$.  Without loss of generality we can then take
$y_0^2=x^2(x-U)$.
\lfm{2.} The constants $a,\ b,\ c$ in \torusc\ are holomorphic in $U$
and the various coupling constants.  This guarantees that $\tau$ is
holomorphic in them.
\lfm{3.} The curve \torusc\ must be compatible with all the global
symmetries of the theory including those which are explicitly broken by
the coupling constants or the anomaly.
\lfm{4.} In various limits (e.g.\ as some mass goes to zero or infinity)
we should recover the curves of other models.
\lfm{5.} The curve should have the correct monodromies around the
singular points.

\noindent
In the next section we will give several examples demonstrating how
these constraints can be used to determine $\tau$.

As in \ntwo, we can add a superpotential $W_{tree}$ to the microscopic
theory and lift the flat directions.  Adding it in the low energy theory to
\lowengw, we find that
\eqn\lowengwa{\eqalign{
&\sum_{l=1}^{k_i} c_l^{(i)} \tilde E_l^{(i)}  E_l^{(i)} + {\partial
W_{tree}\over \partial U}=0 \cr
&(U-U_i) E_l^{(i)} =(U-U_i) \tilde E_l^{(i)} =0 \cr
&\sum_l  |E_l^{(i)}|^2 - |\tilde  E_l^{(i)}|^2=0 .}}
Therefore, for nonzero ${\partial W_{tree}\over \partial U}$, at least
one of the light charged field pairs $E_l^{(i)}$ and $\tilde E_l^{(i)}$
acquire expectation values, giving the photon a mass by the Higgs
mechanism.  Since the $ E_l^{(i)}$'s are magnetic monopoles, their
expectation values lead to confinement of the elementary fields.  In
addition, $U$ is ``locked'' at the singularity $U_i$.

In many examples the expectation value of $U{\partial W_{tree}\over
\partial U}$ is closely related, by the Konishi anomaly \cernrev, to
the expectation value of the ``glueball'' superfield $S=-{1\over 32\pi^2}
\Tr W_\alpha ^2$.  $\ev{S}$ is then
related by \lowengwa\ to the expectation values of the monopoles $E_l^{(i)}$:
\eqn\cons{\ev{S} \sim \ev{U\sum_{l=1}^{k_i} c_l^{(i)} \tilde E_l^{(i)}
E_l^{(i)}}.}
Therefore, $\ev{S}= \ev{\lambda\lambda}/32\pi^2 \not= 0$ only when
monopoles condense; i.e.\ when the theory confines.  There are however
situations where the expectation value of $U{\partial W_{tree} \over
\partial U}$ is not simply related to $\ev{S}$ by the Konishi anomaly.
This is the case in the last example in this paper; but there too we
find that $\ev{S}$ is proportional to the monopole bilinear.  It is
therefore tempting to speculate that perhaps $\ev{S}$ can always be used
as a local order parameter for confinement.  There is also the
possibility that $\ev{U{\partial W_{tree}\over \partial U}}=
\ev{S}= 0$ with nonzero monopole expectation values either because
the terms in \lowengwa\ sum to zero or because $\ev{U}=0$.
Both possibilities indeed occur in some of the theories of \ntwoii.
However, in those
models there is no distinction between the Higgs phase and the
confining phase and therefore we do not know if there is a counter
example to this speculation.

The theory with $W_{tree}=0$ has a Coulomb phase where there is a
photon and the field $U$ whose expectation value can be varied,
serving as a moduli space coordinate for the Coulomb phase.  When
$W_{tree}$ is turned on, the theory passes to the confining phase
where $U$ is locked at one of the singular values $U_i$, the photon is
lifted, and charges are confined.  The variable coordinates of the
confining phase are the parameters in $W_{tree}$.  That phase has two
equivalent descriptions.  The first is the previous one involving the
fields of the Coulomb phase and the light monopoles; this description
has a smooth $W_{tree}\rightarrow 0$ limit.  The other is in terms of
the gauge invariant fields of the confining phase; it is not valid at
the transition point.

Techniques for finding a gauge invariant description of the confining
phase were discussed in \iin .  For the
general situation where there can be a Coulomb phase and the confining
phase and the Higgs phase can be distinct, these techniques only probe
the confining phase and thus lead to incomplete results.  For
example, they are bound to give a constraint which fixes $U$
to be at one of the singular values $U_i$, which is correct for the
confining phase but misses the $U$ moduli space of the Coulomb phase.
In the examples in the next section we will show that this description
of the confining phase is connected properly to the Coulomb phase.

To summarize, it is generic to find points on the moduli space with
massless magnetic monopoles. Furthermore, appropriate perturbations in
the microscopic theory lead to condensation of these monopoles and
confinement.  These observations in \ntwo\ are thus more
general and are not limited to theories with $N=2$ supersymmetry.
However, unlike the $N=2$ theories, our discussion only concerns $\tau$
and does not determine the dyon masses, nor does it determine the Kahler
potential and the metric on the moduli space.

\newsec{Examples}

\subsec{$SU(2)$ with a triplet $\phi$}

This is the example considered in \ntwoi .  The gauge
singlet superfield $U=\Tr \phi ^2$ is a coordinate on the quantum moduli
space. As discussed in \ntwoii , the curve \torusc\ which gives $\tau$
in the Coulomb phase is
\eqn\niicrv{y^2=x^3-x^2U+{1\over 4}\Lambda ^4x.}
Turning on a superpotential $W_{tree}=m U$ gives the field $\phi$
a mass.  Below the scale $m$, $\phi$ can be integrated out and the
theory becomes $N=1$, $SU(2)$ Yang-Mills theory.  The Konishi anomaly
\cernrev\ gives $\ev{\phi {\partial W_{tree}\over \phi}}=4\ev{S}$ where
the factor of 4 arises from the index of the adjoint representation.
This gives $m\ev{U}=\pm 2\Lambda _d^3$ where $\Lambda _d$ is the scale
of the low energy $SU(2)$ theory and we used the known $SU(2)$ gaugino
condensation result $\ev{S}=\pm \Lambda _d^3$.  This constraint on the
value of $U$ is indeed appropriate for the confining branch: these are
the values $U_i=\pm \Lambda ^2$, of $U$ where \niicrv\ is singular.
The matching relation between the scale $\Lambda_d$ and the
scale $\Lambda$ of the high-energy theory which includes $\phi$ is thus
$\Lambda _d^6={1\over 4}m^2\Lambda ^4$, where the 4 reflects the above
normalization conventions, chosen to agree with those of
\refs{\ntwoii,\iin}.

The constraint on $U$ on the confining branch can be nicely seen by
starting from the superpotential for the massive field $S$ of $SU(2)$
Yang-Mills theory \cerne
\eqn\gcw{W_d=S\left(\log\left({m^2\Lambda ^4\over 4S^2}\right)+2\right),}
where we used the above relation between $\Lambda _d$ and $\Lambda$.  As
discussed in \iin, the superpotential $W=W_u+mU$ of the theory which
includes $\phi$ can be determined by ``integrating in'' $\phi$, using
$W_u=[W_d-mU]_{\langle m\rangle}$ where $\langle m\rangle $ solves
${\partial\over \partial m}(W_d-mU)=0$; this is the inverse of
integrating out $\phi$: $W_d=[W_u+mU]_{\ev{U}}$.  The result for
$W=W_u+mU$ is then
\eqn\ntwowiths{W=S\log ({\Lambda^4\over U^2})+mU.}
The equations of motion give
\eqn\ntwowithse{\eqalign{U&=\pm  \Lambda^2 \cr
S&= \half m U.}}
Note that, as in the discussion after \lowengwa, $\ev{S}$ is an order
parameter for confinement in this theory.

These results are indeed correct for $m\neq 0$.  On the other hand,
the $m\rightarrow 0$ limit of \ntwowiths\ continues to give the
constraint $U=\pm \Lambda^2$ of the confining branch, whereas $U$ is
unconstrained on the Coulomb branch.  We conclude that \ntwowiths\ is
valid only on the confining branch.  It has already been observed in
\ntwo\ that often effective Lagrangians are valid only in patches and
not everywhere on the space of parameters and moduli.

\subsec{$SU(2)_1\times SU(2)_2$ with two fields $Q_f$ ($f=1,2$) in the
$({\bf 2,2})$}

The gauge singlets are $M_{fg}=Q_f\cdot Q_g \equiv \half
Q_{f, c_1c_2}Q_{g, d_1d_2}\epsilon ^{c_1d_1}\epsilon ^{c_2d_2}$ in the
$\bf 3$ of the global $SU(2)_F$ flavor symmetry.  The theory has a 3
complex dimensional moduli space of vacua labeled by $M_{fg}$; these
vacua remain exactly degenerate in the quantum theory because the
symmetries do not allow any superpotential to be generated.
Classically, when $U\equiv\det M_{fg} \not= 0$ the $SU(2)_1\times SU(2)_2$
gauge group is broken by the Higgs mechanism down to a $U(1)_D\subset
SU(2)_D$, where $SU(2)_D$ is a diagonally embedded $SU(2)$ subgroup.
Therefore, the theory is in the Coulomb phase.

By the $SU(2)_F$ flavor symmetry, the function $\tau$ of the Coulomb
phase is $\tau (U, \Lambda _1^4, \Lambda _2^4)$, where
the scales $\Lambda _1$ and $\Lambda _2$ refer to $SU(2)_1$ and
$SU(2)_2$ and the exponents are those of instanton contributions,
$e^{-8\pi ^2/g_i^2(\mu )}=(\Lambda _i/\mu )^4$.  Consider the
limit of large $U$.  Taking $M_{11}$ large, the gauge symmetry is
broken to $SU(2)_D$, with $Q_2$ decomposing into a singlet and a
triplet $\phi _D$, and there is also the singlet $M_{11}$.
The low energy theory is then similar to the
previous example with the light field $U_D=\Tr \phi _D^2=2U/M_{11}$,
where the 2 is from the trace, and with a scale $\Lambda _D$ related to
the $\Lambda _i$ by\foot{The threshold factor 16 can be determined
by giving $Q_2$ a mass $m$.
In the resulting low energy theory \iin ,
$\tilde \Lambda _1^5\tilde \Lambda _2^5/M_{11}^2=\tilde \Lambda _D^6$.
As usual for matter in the fundamental,
$\tilde \Lambda _i^5=m\Lambda _i^4$.  But our matching condition
for adjoint fields gives
$\tilde \Lambda _D^6=\tilde m^2\Lambda _D^4/4$ where $\tilde m
=mM_{22}/U_D=m/2$.} $\Lambda _D^4=16\Lambda _1^4\Lambda
_2^4/M_{11}^2$.  In this limit, $\tau$ is therefore given by the curve
\niicrv\ for $U_D$ and $\Lambda _D$; this gives
\eqn\largeucrv{y^2=x^3-x^2U+x\Lambda _1^4\Lambda _2^4 \qquad\hbox{for
large $U$},}
where we used the above relations and we rescaled $x\rightarrow
2x/M_{11}$ and $y\rightarrow (2/M_{11})^{3/2}y$.
This curve gives
\eqn\qqweak{\tau\approx {i\over \pi}\log \left({U^2\over \Lambda_1^4
\Lambda_2^4}\right)\qquad\hbox{for large $U$};}
so taking $U\rightarrow e^{2\pi i}U$ gives a monodromy of
$\CM_{\infty}=T^{-4}$.  This monodromy means that there must be (at
least two) strong coupling singular points where monopoles become
massless.

Using the symmetries, including the
${\bf Z}_2$ symmetry which exchanges the two gauge groups, the exact
coefficient $a$ in \torusc\ must be $a=-U+\alpha (\Lambda_1^4+\Lambda
_2^4)$, for some constant $\alpha$.  In addition, agreement with
\largeucrv\ requires $b=\Lambda _1^4\Lambda _2^4$ and $c=0$ to be exact
expressions.  Note that this curve is always singular when either
$\Lambda _1$ or $\Lambda _2$ vanishes.  This reflects the fact that
the low energy photon decouples, and hence $\tau
\rightarrow i\infty$, in these limits.
To determine the remaining parameter $\alpha$, we consider the
limit $\Lambda _2\gg \Lambda _1$.  In this limit the
theory is approximately an $SU(2)_1$ gauge theory with three singlets
$M_{fg}$ as well as a field $\tilde \phi$ in the adjoint of $SU(2)_1$.
These fields are related by the constraint of \nati
\eqn\iipf{\pf V=U+\mu ^2\tilde U=\Lambda _2^4,}
where $\mu$ is a dimensionful normalization needed to make
$\tilde\phi$ a canonical field.   $SU(2)_1$
with a field $\tilde\phi$ in the adjoint is singular at $\tilde U=\pm
\Lambda _1^2$.  Using \iipf, $\tau$ should then be singular at $U\approx
\Lambda _2^4\pm \mu ^2\Lambda_1^2$ in the $\Lambda_2 \gg
\Lambda_1$ limit.  On the other hand, the discriminant reveals that
our curve is singular at $U=\alpha (\Lambda_1^4+\Lambda_2^4) \pm
2\Lambda _1^2\Lambda _2^2$.  Comparing, we find that
$\alpha=1$.  To summarize, we have determined that $\tau$ is given
by the curve \eqn\qqxy{y^2=x^3+x^2(-U+\Lambda _1^4+\Lambda _2^4)+
\Lambda _1^4\Lambda_2^4x.}

In addition to the weak coupling singularity \qqweak, this $\tau$ is
singular for two values of $\det M\equiv U$ in the strong coupling
region: $U_i=(\Lambda_1^2 \pm\Lambda_2^2)^2$.  Note that in terms of the
moduli space of vacua given by the expectation values of the $M_{fg}$,
these are singular (non-compact) submanifolds rather than singular
points.  The order of the zero of the discriminant at these
singularities implies that they both have monodromy conjugate to $T$.
There is thus a single massless field on each of the two $U_i$
submanifolds.  Their charges are $(n_m, n_e)$=$(1,0)$ for one of the
singular spaces and $(1,1)$ for the other, much as in \ntwoi .

Near either of the two strong coupling singular submanifolds of $\det
M=U_i=(\Lambda _1^2\pm \Lambda _2^2)^2$, the low energy theory is
approximately described by the effective superpotential
\eqn\weffxpm{W^{(i)} \approx c^{(i)}(\det M-U_i)\tilde E^{(i)}E^{(i)}
+\Tr\ mM.}
For $m\neq 0$, the equations of motion give $M=\epsilon \sqrt{\det m
U_i}m^{-1}$ and $ c^{(i)}\mmp= -\epsilon \sqrt{{\det m\over U_i} }$,
where $\epsilon =\pm 1$.  So each singular submanifold $U_i$ gives two
vacua in the fully massive theory, for a total of four vacua.  In the
limit of large $m$ the $Q_f$ can be integrated out and these four vacua
go over to the four vacua of the low energy
$SU(2)\times SU(2)$ Yang-Mills theory. As in the previous example, the
condensation of the monopoles leads to confinement.  Even though the
theory has matter fields in the fundamental representation of each
$SU(2)$ factor, the Higgs and confining phases are different.  All the
matter fields are invariant under the diagonal ${\bf Z}_2$ subgroup of
the centers of the two $SU(2)$ factors.  Therefore, Wilson loops in
representations which are affected by this ${\bf Z}_2$, say $({\bf
2,1})$, should exhibit area law in the confining phase.

The singular values $U_i$ of $U$ can also be determined directly by
giving the $Q_f$ masses and considering the confining phase, where $U$
is automatically locked at the $U_i$.  Starting from the low-energy
$SU(2)_1\times SU(2)_2$ Yang-Mills theory, the $Q_f$ can be integrated
in using the technique of \iin .  This is essentially the reverse of the
discussion of the previous paragraph.  The result is that the confining
branch is described by
\eqn\wqqss{W=S_1\log \left({\Lambda_1^4(S_1+S_2)^2 \over S_1^2\det M}
\right) + S_2\log \left({\Lambda_2^4(S_1+S_2)^2 \over S_2^2\det M}
\right)+\Tr\ mM.}
Upon integrating out the $S_s$, this gives
\eqn\wqq{W=\Tr\ mM \qquad\hbox{with}\qquad \det M=U_\pm\equiv
(\Lambda_1^2\pm \Lambda_2^2)^2.}
For example, giving a mass $m_2$ to the matter field $Q_2$ and
integrating it out subject to \wqq\ gives the effective Lagrangian
found in \iin\ for the remaining light field $M_{11}$.  The vacua have
$\ev{S_s}=\epsilon _s\sqrt{\det m\Lambda _s^4}$, with $\epsilon
_1\epsilon _2=\pm 1$ for $\ev{U}=(\Lambda _1^2\pm \Lambda _2^2)^2$.
Again we see that these are order parameters for the confinement which
occurs for nonzero $\det m$.

In the confining phase \wqq, as discussed in the previous section, $U$
is indeed locked at the singular values $U_i$ of the Coulomb phase
determined from the curve \qqxy.  In fact, having determined the
$U_i$ by thus analyzing the confining phase, we could have bypassed
some of the previous detailed analysis of the curve by knowing the
values of the $U_i$.

\subsec{$SU(2)$ with two triplets $\phi_f$ ($f=1,2$)}

The classical moduli space is parametrized by the expectation values
of the gauge singlet fields $M_{fg}=\Tr (\phi _f\phi _g)$, which
transform in the $\bf 3$ of the global $SU(2)_F$ flavor symmetry.  For
generic values of $M$ the $SU(2)$ gauge symmetry is completely broken
and the theory is in the Higgs phase.  On the submanifold with $\det
M=0$ there is an unbroken $U(1)$ gauge symmetry and thus a light
photon -- this subspace is in the Coulomb phase.

This vacuum degeneracy cannot be lifted quantum mechanically.  The
reason is that the only invariant superpotential for the
light fields is proportional to $\det M/\Lambda$.  This does not have a
proper semi-classical limit ($\Lambda \rightarrow 0$) and therefore
cannot be generated to lift the degeneracy of the Higgs phase or the
Coulomb phase subspace.  Below we will
argue that such a superpotential does give a proper description of the
confining phase.

Now consider adding a tree level superpotential $W_{tree}=\Tr\ mM$.
For $\det m\neq 0$ it gives both $\phi_f$ a mass and the low energy
theory is an $N=1$ pure gauge $SU(2)$ theory, which is known to
confine.  For $m\neq 0$ but $\det m=0$ only one of the matter fields
gets a mass and the low energy theory is an $N=2$ pure gauge theory,
which is in the Coulomb phase \ntwoi.  We see that the theory can be
in all three of the different possible phases.  Note that, because the
theory contains no fields in the fundamental representation of the
gauge group, the confining and Higgs phases are distinct \fund\ -- the
Wilson loop has an area law in the confining phase and a perimeter law
in the Higgs phase.

The confining phase, using the technique of \iin,
is described by
\eqn\sppwu{W= S\left[\log \left({\Lambda^2S^2\over \det M^2}\right)
-2\right]+\Tr\ mM.}
Here we used the matching relation $\Lambda _d^6=\det m^2\Lambda^2/16$
between the scale $\Lambda$ of the high-energy theory which includes
the two $\phi_i$ and the scale $\Lambda_d$ of the low-energy $N=1$
pure gauge $SU(2)$ theory (this relation is the one discussed for
$SU(2)$ with a single adjoint, applied for each eigenvalue of $m$).
Integrating out the field $S$ by its equation of motion gives $\ev
S=\pm \Lambda ^{-1}\ev{\det M}$ and the superpotential
\eqn\wppm{W=\mp 2{\det M\over \Lambda}+\Tr\ mM.}
Upon integrating out $M$, we obtain
\eqn\MSevs{\ev M=\pm \half (\Lambda \det m)m^{-1}\qquad \hbox{and}\qquad
\ev S=\pm {1\over 4}\Lambda\det m .}
These are the correct expectation values for the confining phase.

However if we take $m=\pmatrix{0&0\cr 0&m_{22}}$, giving mass to
$\phi_2$ only, the theory is actually in the Coulomb phase.  Nevertheless,
\MSevs\ gives $\Tr \phi _1^2=\pm \half m_{22}\Lambda=\pm \Lambda_I^2$
where $\Lambda _I$, the scale of the theory with $\phi_2$ integrated
out, is related to $\Lambda$ by the matching relation for adjoint
fields.  We see, once again, that continuing from the confining
phase gives only the singular values $U_i$, missing the Coulomb
phase moduli space of $U$.
In this phase, as in \niicrv , the curve which determines $\tau$ is
\eqn\curvcphph{y^2=x^2(x-M_{11}) + {1 \over 16} m_{22}^2\Lambda^2 x.}
It is singular as $m_{22} \rightarrow 0$.  This is because there are
massless charged ``electrons'' on the Coulomb branch for $m=0$.  The low
energy gauge coupling is therefore
renormalized to zero in the infra-red.  Hence
$\tau \rightarrow i \infty$ and the curve must be singular.

We can now see the transitions between the various phases.  For $m=0$
the $\det M=0$ submanifold of the moduli space is in the Coulomb phase
and there are charged ``electrons'' on this submanifold.  The
transition to the Higgs phase is characterized by the expectation
values of these electrons.  For $m\not=0$ but $\det m=0$ the theory
has only a Coulomb phase while for $\det m \not=0$ there is only a
confining phase.  The transition between them takes place by monopole
condensation at $M= \pm \half (\Lambda \det m) m^{-1}$.  Again, the
expectation value \MSevs\ of $S$ is a local order parameter for the
$\det m\not=0$ confining phase, as expected from the discussion
following \lowengwa.

A direct transition from the Higgs phase to the confining phase
happens at $M=0$ for $m=0$.  The simplest interpretation of the
physics at that point is that the elementary gauge bosons of the
underlying $SU(2)$ theory are massless there.  Hence, this theory must
be at a non-trivial fixed point of the renormalization group.
Examples of such fixed points have already appeared in \nati .

\subsec{$SU(2)$ with two doublets $Q_f$ ($f=1,2$) and a triplet $\phi$}

The basic gauge singlets are $X=Q_1Q_2$, $U=\Tr \phi^2$, and $\vec
Z={\sqrt{2}\over 2}Q_f \phi Q_g\vec \sigma ^{fg}$ transforming in the
$\bf 3$ of the global $SU(2)_F$ flavor symmetry.  Classically, there
is a moduli space of inequivalent vacua given by the expectation
values of these fields subject to the classical constraint $\vec
Z^2=X^2U$. In the
quantum theory the above five fields are independent with a
superpotential which is determined by the
symmetries to be of the form
\eqn\wpqqf{W_u=-{XU^2\over \Lambda ^3}f\left( t={\vec Z^2 \over
X^2U}\right).}
We will determine the function $f(t)$ shortly.  In order to reproduce
the correct asymptotic behavior of the moduli space, the function
$f(t)$ in \wpqqf\ must have a double zero at $t=1$;
$f(1)=f^\prime(1)=0$.  There is then a quantum moduli space of vacua
characterized by $t=1$; i.e.\ by $\vec Z^2=X^2U$.  For generic
expectation values of the fields the $SU(2)$ gauge symmetry is
completely broken and the theory is in the Higgs phase.  Note that,
since there is no distinction between the Higgs and the confining
phase for this theory, we expect \wpqqf\ to describe the theory
everywhere away from the Coulomb phase.

The quantum moduli space is singular on the subspace $X=\vec Z=0$ for
any $U$.  On that complex line there is an unbroken $U(1)$ gauge
symmetry and the theory is in the Coulomb phase.  The physical reason
for the singularity is that $X$ and the $\vec Z$ are not the
correct light fields on this submanifold: in the Coulomb phase the
elementary charged fields $Q_f$ are massless.  As in the previous
example, these fields get expectation values off the Coulomb
submanifold, leading to a transition to the Higgs phase.

Consider turning on a tree level superpotential $W_{tree}=m_Q X+\vec
\lambda \cdot \vec Z$.  As long as either $m_Q$ or $\vec \lambda$ is
nonzero, this superpotential fixes the theory to lie on the Coulomb
phase submanifold.  For $\vec \lambda ^2=1$, the theory is $N=2$
supersymmetric and was analyzed in \ntwoii.  It was found there that
the function $\tau$ is described by the curve
\eqn\niitorus{y^2=x^2(x-U)+{1\over 4}m_Q\Lambda ^3x-{1\over 64}\Lambda
^6\qquad\hbox{for}\qquad \vec \lambda ^2=1.}  This can be immediately
generalized to arbitrary $\vec \lambda^2$ (it only depends on $\vec
\lambda ^2$ by the $SU(2)_F$ flavor symmetry).  The theory has a
global $U(1)_Q\times U(1)_\phi \times U(1)_R$ symmetry with the
charges $U:(0,2,0)$, $m_Q:(-2,0,2)$, $\vec \lambda^2:(-4,-2,4)$, and
$\Lambda ^3:(2,4,-2)$ (using the anomaly as in \iin).  The terms in
\niitorus\ should thus have charges $(0,6,0)$ and hence
\eqn\abcis{y^2=x^3-x^2U+{1\over 4} m_Q\Lambda ^3x -{1\over 64}\vec
\lambda ^2\Lambda ^6.}
This curve is singular when $m_Q=\vec \lambda=0$.  In this case there
are massless ``electrons'' in the Coulomb phase and they renormalize
the electric charge to zero in the infra-red.  Hence $\tau=i\infty$
and the curve is singular.

In the Higgs phase, adding $W_{tree}=m_QX+\vec \lambda \cdot \vec Z$,
the theory is described by the superpotential $W=-{XU^2\over \Lambda
^3}f(t)+W_{tree}$.  On the other hand, because there are matter fields
in the fundamental representation of the gauge group, there is no
phase boundary separating the Higgs phase and the confining phase.
Therefore, we can approach the Higgs phase from the Coulomb line via the
confining phase.  The
expectation value of the field $U$ should then be at the singular
values $U_i$ obtained from the curve \abcis.  Therefore, the equations
of motion obtained from this Higgs phase superpotential must fix $U$
at the singular values $U_i$ of the curve \abcis.  Indeed, upon
integrating out the massive fields $X$ and $\vec Z$, the Higgs phase
superpotential becomes $W=0$ with the constraints
\eqn\conuf{U^2(f-2tf')=m_Q\Lambda ^3\equiv 4b \qquad \hbox{and}\qquad
4U^3f'^2t=\vec \lambda ^2\Lambda ^6\equiv - 64c.}
Eliminating $t$, these equations fix $U$ to particular values $U_i$.  On
the other hand, using \abcis\ and \singabc, the singular values $U_i$
of $U$ satisfy
\eqn\discubc{-4U^3c-U^2b^2+18Ubc+4b^3+27c^2=0.}
The equations \conuf\ and \discubc\ must agree for every value of the
parameters $b$ and $c$.  This gives a differential equation for $f$:
\eqn\fdeq{4f'^2t-(f-2tf')^2-{9\over 2}(f-2tf')f'^2t+ (f-2tf')^3+
{27\over 16}f'^4t^2=0.}
This differential equation has a unique solution subject to the boundary
conditions $f(1)=f'(1)=0$ discussed above.  The unique solution is
\eqn\unisol{f(t)=(1-t)^2.}

The equations \conuf\ with $f=(1-t)^2$ are
\eqn\conyu{U^2(1-t)(1+3t)= m_Q\Lambda^3\qquad\hbox{and}\qquad
16U^3t(1-t)^2=\vec \lambda ^2\Lambda^6 ;} they are equivalent to the
singularity equations \toruss\ with the substitution $2x=U(1-t)$.  Note
that, for $m_Q$ or $\vec \lambda$ nonzero, $t\not=1$; adding $m_Q$ or
$\vec \lambda$ takes the theory off its quantum moduli space.  This
phenomenon has already been observed in \refs{\nati, \iin,\ntwoii}.  For
example, for $\vec\lambda =0$ with $m_Q\neq 0$, \conyu\ gives $U_{1,2}=\pm
\sqrt{m_Q\Lambda ^3}$ with $t=0$ and $U_3=\infty$ with $t=1$ (as
expected classically since the coupling is weak at
$U=\infty $).  For $m_Q=0$
and $\vec \lambda\neq 0$, \conyu\ gives $U_{1,2,3}^3=-{27\over
256}\vec\lambda ^2\Lambda ^6$ with $t=-{1\over 3}$. The monodromy
around each of the singular values $U_i$ of $U$ is conjugate to $T$
and, thus, a single light field is present at each $U_i$.

The superpotential \wpqqf\ with $f(t)=(1-t)^2$ can also be obtained by
considering the the confining phase.  Consider turning on the
tree-level superpotential $W_{tree}=m_\phi U+\vec \lambda \cdot \vec
Z$ and integrating out $\phi$.  Doing so at tree level, we obtain
$W_{tree,d}=-{\vec \lambda ^2\over 4 m_\phi}X^2$.  Including the gauge
dynamics, the full superpotential in the theory with $\phi$ integrated
out is uniquely determined, as in \iin, to be simply
$[W_u+m_\phi U+\vec \lambda \cdot \vec Z]_{\ev U, \ev{\vec
Z}}={m_\phi ^2\Lambda ^3\over 4X}-{\vec \lambda ^2\over 4 m_\phi}X^2$.  Any
modification of this result can be ruled out using the symmetries,
holomorphy, and the behavior in various limits. Thus
$W_u=[{ m_\phi ^2\Lambda ^3\over 4 X}-{\vec \lambda^2\over
4m_{\phi}}X^2-m_\phi U-\vec \lambda \cdot Z]_{\ev {m_\phi}, \ev {\vec
\lambda}}$ and hence
\eqn\wxyzi{W_u=-{XU^2\over \Lambda ^3}\left(1-{\vec Z^2 \over
X^2U}\right)^2,}
reproducing the result $f=(1-t)^2$.  Having obtained this result, we
can add $W_{tree}=m_QX+\vec \lambda \cdot \vec Z$, use the equations
of motion \conyu\ to find the singular values $U_i$ of $U$, and thus
reproduce the answer obtained from the curve \abcis\ of the Coulomb
branch of the theory.  Note that it is possible to explore the entire
Higgs phase moduli space of $t=1$ from the confining phase expectation
values obtained by perturbing the theory by
$W_{tree}=m_\phi U+ m_QX+\vec\lambda\cdot \vec Z$.  In particular,
taking $m_Q, \, m_\phi,\, \vec \lambda \rightarrow 0$ with $\vec
\alpha= \vec \lambda /m_Q,\, \tilde m_\phi=m_\phi/m_Q$ fixed, we find
$\ev{1-t} \approx {\Lambda^3 m_Q (\vec \alpha ^2)^2 /4}\rightarrow 0$,
$\ev U \approx {1/ \vec \alpha^2}$, $\ev X \approx {2 \tilde m_\phi /
\vec \alpha^2}$, and $\ev{\vec Z} \approx -2\vec \alpha \tilde m_\phi
/(\vec \alpha^2)^2$.  By adjusting $\vec \alpha$ and $\tilde m_\phi$ in
this limit, these expectation values explore the entire $t=1$ moduli
space.

The superpotential with the massive field $S$ integrated in is
\eqn\wxyzs{W_u=S\left[\log \left({\Lambda ^3S\over XU^2(1-t)^2}\right)
-1\right];}
this gives \wxyzi\ upon integrating out $S$ by its equation of motion
$\ev S=XU^2(1-t)^2\Lambda ^{-3}$.
Perturbing by $W_{tree}=m_\phi U+ m_QX+\vec\lambda \cdot \vec Z$ and
integrating out $X$ and $\vec Z$ gives
\eqn\soncb{W=S\log\left({m_Q\Lambda ^3\over U^2(1-t_0)(1+3t_0)}\right)
+m_\phi U\qquad\hbox{with}\qquad t_0^{-1}(1+3t_0)^2=
{16m_Q^2\over \vec \lambda ^2U},}
which is equivalent to \conyu\ upon integrating out $S$.  Integrating
out $U$ gives $\ev{S}$ which is proportional to $m_\phi$ showing, again,
that it is an order parameter for confinement in this theory.

There is one point on the moduli space which we have not yet discussed.
For $m_Q=m_\phi=\vec \lambda=0$ there is a ground state with $X=U=\vec
Z=0$.  This point can be approached from several directions.  Moving in
{}from nonzero $U$ with $X=\vec Z=0$ we conclude that there must be a
photon and massless charged fields there.  If we instead set
$m_\phi=0$ and take $m_Q$ and $\vec \lambda$ to zero in various
ratios we find at $U=0$ a photon and either two or three monopoles.
We expect that the  resolution is that the correct degrees of freedom at
that point are actually the elementary quarks and gluons and that this is
another scale invariant theory.

\centerline{{\bf Acknowledgements}}

We are particularly indebted to E. Witten for very useful discussions.
It is also a pleasure to thank T. Banks, R. Leigh, M.R. Plesser,
and S. Shenker for interesting conversations.
We also acknowledge the Aspen Center for
Physics, where some of this work was conducted.  This work was supported
in part by DOE grant \#DE-FG05-90ER40559.

\listrefs

\end